\documentclass[prd,aps,showpacs,floats]{revtex4}
\usepackage{amsmath,amssymb}
\usepackage{graphicx}
\def\lsim{\raise0.3ex\hbox{$\;<$\kern-0.75em\raise-1.1ex
\hbox{$\sim\;$}}}
\def\gsim{\raise0.3ex\hbox{$\;>$\kern-0.75em\raise-1.1ex
\hbox{$\sim\;$}}}

\DeclareMathAlphabet{\mathsc}{OT1}{cmr}{m}{sc}

\newcommand{\Atm}  {\mathsc{atm}}

\newcommand{\Lsnd} {\mathsc{lsnd}}
\newcommand{\Ch}   {\mathsc{ch}}

\newcommand{\dms}{\Delta m^2_\odot}
\newcommand{\dma}{\Delta m^2_\Atm}
\newcommand{\dml}{\Delta m^2_\Lsnd}
\newcommand{\sql}{\sin^2 2\theta_\Lsnd}

\newcommand{\myodot}{{\mathchoice
    {\raisebox{\depth}{$\displaystyle\odot$}}
    {\raisebox{\depth}{$\textstyle\odot$}}
    {\raisebox{\depth}{$\scriptscriptstyle\odot$}}
    {\raisebox{\depth}{$\scriptscriptstyle\odot$}}
    }}

\begin{document}
\overfullrule 0pt
\date{\today}

\title{\
\vglue -2.0cm
{\small \hfill IFT-P.007/2003}\\
\vglue -0.1cm
{\small \hfill IFUSP-DFN/03-079}\\
\vglue -0.1cm
{\small \hfill NSF-ITP-03-09}\\
\vglue -0.1cm
{\small \hfill hep-ph/0302039}\\
\vglue 1.0cm
Probing the LSND mass scale and four neutrino scenarios with a 
neutrino telescope}

\author{H. Nunokawa$^{1,2}$}\email{nunokawa@ift.unesp.br}
\author{O. L. G. Peres$^{2,3}$}\email{orlando@ifi.unicamp.br} 
\author{R. Zukanovich Funchal$^{2,4}$}\email{zukanov@if.usp.br} 

\affiliation{\\ \\
 $^1$ Instituto de F\'{\i}sica Te\'orica,Universidade Estadual
 Paulista, Rua Pamplona 145, 01405-900 S\~ao Paulo, Brazil \\ 
 $^2$ Kavli Institute for Theoretical Physics, University of California,
 Santa Barbara, CA 93106, USA\\ 
 $^3$ Instituto de F\'{\i}sica Gleb Wataghin,
 Universidade Estadual de Campinas -- UNICAMP, 13083-970 Campinas,
 Brazil \\ 
 $^4$ Instituto de F\'{\i}sica, Universidade de S\~ao Paulo
 C.\ P.\ 66.318, 05315-970 S\~ao Paulo, Brazil}

\begin{abstract}
 We show in this paper that the observation of the angular
 distribution of upward-going muons and cascade events induced by
 atmospheric neutrinos at the TeV energy scale, which can be performed
 by a kilometer-scale neutrino telescope, such as the IceCube
 detector, can be used to probe a large neutrino mass splitting,
 $\vert \Delta m^2 \vert \sim (0.5-2.0)$~eV$^2$, implied by the LSND
 experiment and discriminate among four neutrino mass schemes. This is
 due to the fact that such a large mass scale can promote non
 negligible $\nu_\mu \to \nu_e,\nu_\tau$/$\bar \nu_\mu \to \bar
 \nu_e,\bar\nu_\tau$ conversions at these energies by the MSW effect
 as well as vacuum oscillation, unlike what is expected if all the
 neutrino mass splittings are small.
\end{abstract}


\maketitle
\thispagestyle{empty}
\section{Introduction}
\label{sec:intro}
 \vglue -0.2 cm Atmospheric and solar neutrino experiments present
 today compelling evidence of neutrino flavor oscillations induced,
 respectively, by the mass squared differences $\vert \dma \vert \sim
 3\times 10^{-3}$~eV$^2$~\cite{Gonzalez-Garcia:2002sm} and $\dms\sim
 7\times 10^{-5}$~eV$^2$~\cite{ntz}. These results have been
 independently supported by the K2K experiment~\cite{k2k}, confirming
 the $\nu_\mu/\bar \nu_\mu$ conversion observed in atmospheric
 neutrino data, and by the KamLAND experiment~\cite{kamland}, recently
 observing $\bar \nu_e$ disappearance compatible with what has been
 seen by the solar neutrino experiments.

 In the standard three neutrino flavor oscillation framework, the
 above results imply neutrino oscillation lengths at energies
 above about 0.3~TeV much greater than the Earth's diameter, strongly
 suppressing $\nu_{\mu} \rightarrow \nu_{e}/\nu_{\tau}$ and
 $\bar{\nu}_{\mu} \rightarrow \bar{\nu}_{e}/\bar{\nu}_{\tau}$
 oscillations. Moreover, the ratio between $\nu_e/\bar \nu_e$ and
 $\nu_\mu/\bar \nu_\mu$ atmospheric fluxes drops as their energy
 increases so that atmospheric neutrino data above 0.5 TeV are
 expected to be basically composed of $\nu_{\mu}$ and
 $\bar{\nu}_{\mu}$~\cite{review-atm}.  The so called prompt
 atmospheric neutrinos coming from the decay of charmed particles as
 well as extra-galactic
 $\nu_e,\bar \nu_e/\nu_\tau, \bar \nu_\tau$ contributions are
 very small compared to the conventional atmospheric
 $\nu_\mu/\bar \nu_\mu$ fluxes (which mainly come from the decay of
 charged pions and kaons) in the 0.5-10 TeV energy
 range~\cite{Hooper:2002qq}.

 However, as we will see, atmospheric $\nu_e,\bar \nu_e/\nu_\tau,\bar
 \nu_\tau$ fluxes in the TeV energy range can be non-negligible
 compared to the $\nu_\mu/\bar \nu_\mu$ if the result of LSND
 experiment~\cite{LSND} which indicates $\bar \nu_{\mu} \rightarrow
 \bar \nu_e$ transition generated by neutrino oscillations with a
 large mass scale, $\vert \dml \vert \sim (0.5-2.0)$ eV$^2$, turns out
 to be correct.  The LSND result, which will be very soon tested by
 the MiniBooNE experiment~\cite{miniboone}, if confirmed, can be
 interpreted in combination with the atmospheric and solar neutrino
 data as a signature of an additional neutrino, $\nu_s$, sterile in
 nature.  The minimum scheme necessary to explain at same time all
 three indications of neutrino oscillations involves four neutrino
 mass eigenstates~\cite{giunti}.  We are aware that current analyses
 performed with the entire set of neutrino data from accelerators,
 reactors, solar and atmospheric neutrinos seem to disfavor four
 neutrino mass scenarios~\cite{dis,strumia}.  See, however,
 Refs.~\cite{critica} which casted doubt on conclusions drawn in
 Refs.~\cite{dis,strumia}. More importantly, no experiment
 until now has confirmed or ruled out the LSND result and one should
 wait for results from MiniBooNE.

 We will show in this letter that in the framework of four neutrino
 mass schemes, atmospheric neutrino data at the TeV energy scale may
 contain a non negligible amount of $\nu_e$ or $\bar{\nu}_e$, due to
 the interplay of the Earth matter with the LSND mass scale through
 the so called MSW effect~\cite{MSW}, and of $\nu_\tau$ and
 $\bar{\nu}_\tau$, due to almost vacuum oscillations induced by this
 same mass scale.  To our knowledge it has been never mentioned before
 in the literature that the LSND mass splitting could give rise to
 resonant $\nu_\mu \to \nu_e$ or $\bar{\nu}_\mu \to \bar{\nu}_e$
 conversion for atmospheric neutrinos.  These $\nu_e,\bar
 \nu_e/\nu_\tau,\bar \nu_\tau$ events can be observed by a large
 neutrino telescope, such as the proposed IceCube
 detector~\cite{icecube}, by measuring the angular distribution of
 upward-going muons and cascade events.  This is because the zenithal
 dependence of upward-going muons as well as of cascade events will be
 modified by these new contributions with respect to what would be
 expected for the standard three neutrino oscillation framework.  The
 zenithal distortions can not only discriminate among the different
 types of four neutrino mass schemes, but also they are sensitive to
 the sign of the LSND mass splitting, opening a new window to test
 four neutrino mass scenarios.

\vglue -0.5 cm
\section{Framework of four neutrino mixing schemes}
\begin{figure}
\centering\leavevmode
\vglue -1.5cm
\hglue 1.0cm
\includegraphics[width=11.cm]{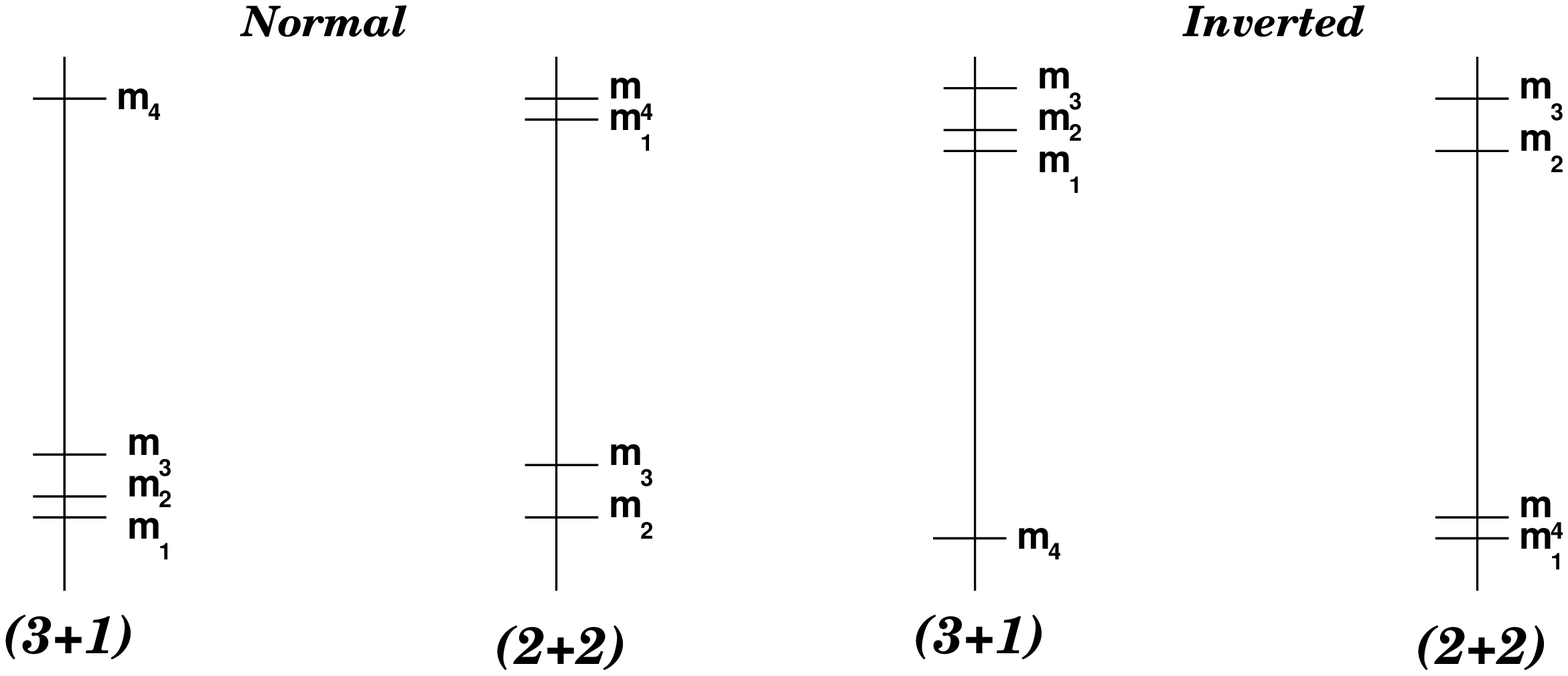}
\caption{Four neutrino mass schemes used in this paper.}  
\label{fig0}
\end{figure}

\vglue -0.2 cm 
 We will work in the minimal model structure needed to describe the three 
 evidence for neutrino oscillations (atmospheric, solar and LSND): a four 
 neutrino mass scheme~\cite{giunti}.  We will adopt the same parametrization 
 and some useful notations used in Refs~\cite{dis,valle-4nu}.  For what 
 concerns us here, the sign of $\Delta m^2_{\myodot}$ and of $\dma$ are 
 irrelevant, only the sign of $\dml$ will be significant.  We can then
 classify the possible four neutrino mass schemes into two categories: 
 (a) {\em (3+1)}, when a single neutrino mass state is separated from the 
 other neutrino mass states by a large mass splitting ($\vert \dml \vert$); 
 and (b) {\em (2+2)}, when two pairs of neutrino mass states are separated 
 by a gap given by the $\vert \dml \vert$ scale. Moreover, when $\nu_4$ is a 
 heavy (light) state we will call it normal (inverted) mass hierarchy.  
 We show in Fig.~\ref{fig0}, the {\it (2+2)} and {\it (3+1)} mass schemes 
 for both, normal and inverted, mass hierarchies which will be considered in 
 this  paper.

 The evolution equations for neutrinos in the flavor basis are
\begin{equation}
\imath \displaystyle \frac{d}{dr} \nu_f =
\left[ U \displaystyle \frac{M^2}{2E_\nu} U^T + A\right] \nu_f, 
\label{evol}
\end{equation}
 where $\nu_f=(\nu_{e}\nu_{\mu}\nu_{\tau}\nu_{s})^{T}$ and $E_\nu$ is
 the neutrino energy.  We define $\dml=\Delta
 m^2_{42}$ and $\dma=\Delta m^2_{32}$, with
 $\Delta m^2_{ij}=m^2_{i}-m^2_{j}$, for $i,j=1,2,3,4$. 
 The diagonalized neutrino mass squared matrix is 
\begin{equation}
M^2 = \rm{diag}(0,-\dml
+\Delta m^2_{\myodot}, \dma+\Delta
 m^2_{\myodot}-\dml,\Delta m^2_{\myodot}),
\end{equation}
with $\Delta
 m^2_{\myodot}=\Delta m^2_{41}$ for $\it{(2+2)}$ cases and
\begin{equation}
M^2=\rm{diag}(0,\Delta m^2_{\myodot},\Delta m^2_{\myodot}+\dma, 
\dml+\Delta m^2_{\myodot}),
\end{equation}
 with $\Delta m^2_{\myodot}=\Delta m^2_{21}$ for $\it{(3+1)}$ ones.  
 We use for the $4\times 4$ mixing matrix $U$ the same parametrization 
 as in Ref.~\cite{valle-4nu},
\begin{equation}
U=R_{34}R_{24}R_{23}R_{14}R_{13}R_{12},
\end{equation}
 where $R_{ij}$ are rotation
 matrices in the subspace $(i,j)$ with angle $\theta_{ij}$.  The
 matter potential matrix is 
\begin{equation}
A=\rm{diag}(A_{\rm{cc}},0,0,A_{\rm{nc}}),
\end{equation}
 with $A_{\rm{cc}}=\sqrt{2}G_F \rho Y_e$ and
 $A_{\rm{nc}}=A_{\rm{cc}}(1-Y_e)/(2Y_e)$, where $G_F$ is the Fermi
 constant, $Y_e$ the electron fraction and $\rho$ the matter density.
 For antineutrinos evolution equations similar to Eq.~(\ref{evol})
 exist but with $A \to -A$.  We have ignored all CP violation phases,
 as they do not play any important r\^ole in our study.

 In order to make clear the connection among mixing parameters and  
 experimental observations, we will introduce some useful definitions.
 Let us  first define the quantities $d_\alpha$ ($\alpha=e,\mu,\tau,s$) as
\begin{equation} \label{eq:defd}
d_\alpha =
\left[\begin{array}{lr}
|U_{\alpha 4}|^2    &  \hskip 1.2cm \text{for {\it (3+1)} mass scheme,} \\
|U_{\alpha 1}|^2 + |U_{\alpha 4}|^2  & \hskip 0.8cm  \text{for {\it
(2+2)} mass scheme,}
\end{array}\right. 
\end{equation}
and $\eta_\alpha$ ($\alpha=e,\mu,\tau,s$) as
\begin{equation} \label{eq:defd4}
\eta_\alpha=
\left[\begin{array}{lr}
 |U_{\alpha 1}|^2+ |U_{\alpha 2}|^2 &
\hskip 1.2cm \text{for {\it (3+1)} mass scheme,}\\
 |U_{\alpha 1}|^2 + |U_{\alpha 4}|^2   &
\hskip 1.2cm \text{for {\it (2+2)} mass scheme,}
\end{array}\right. 
\end{equation}
 which can be related to the neutrino survival and oscillation
 probabilities at short and long-baseline experiments as we will
 see below.
 
 Note that $\eta_s$ measures the sterile admixture in solar neutrino
 oscillations, this parameter describes the fraction of sterile
 neutrino participating in solar oscillations: $\eta_s=0 \;(1)$
 corresponds to pure active (sterile) $\nu_e \to \nu_\tau$ ($\nu_e \to
 \nu_s$) oscillations. Solar neutrino data prefer a small admixture
 of sterile neutrino in the $\dms$ scale, {\it i.e.,} $\eta_s \ll 1$.
 The sterile admixture for atmospheric neutrinos is characterized by
 the parameter $d_s$, in the $\dma$ mass scale, $d_s=1 \; (0)$
 corresponds to pure active (sterile) $\nu_\mu \to \nu_\tau$ ($\nu_\mu
 \to \nu_s$) oscillations. Atmospheric neutrino data prefer a small
 admixture of sterile neutrino in the $\dma$ scale, {\it i.e.,} 
$1-d_s  \ll 1$. 

 We can relate $d_\alpha$ to the neutrino survival probability
 in vacuum for short-baseline disappearance experiments, ignoring ${\cal O}
 (\dma$) corrections, by

\begin{equation} \displaystyle 
 P_{\nu_\alpha\to\nu_\alpha} = 1- A_{\alpha;\alpha} \sin^2
 \left(\frac{\dml}{4E_\nu} L \right)= 1 - 4\, d_\alpha (1-d_\alpha)
\sin^2 \left(\frac{\dml}{4E_\nu} L \right),
\end{equation}
 where $L$ is the baseline distance.  From the null results of
 short-baseline experiments $A_{e;e}$ and $A_{\mu;\mu}$ are limited to
 be very small~\cite{giunti}.

 The conversion probability, relevant for short-baseline appearance
 experiments, can be written as 
\begin{equation}
P_{\nu_\mu\to\nu_e} =  A_{\mu;e} \; \sin^2\left( \frac{\dml}{4E_\nu} L \right),
\label{eq:pld}
\end{equation}
where
\begin{equation} \label{eq:defd3}
 A_{\mu;e} =
\left[\begin{array}{lr}
4|U_{e4}|^2|U_{\mu4}|^2    &  \hskip 1.2cm \text{for {\it (3+1)} mass
scheme,} \\
4|U_{e1}U_{\mu 1}^*+U_{e4}U_{\mu 4}^*|^2 & \hskip 0.8cm  \text{for {\it
(2+2)} mass scheme.}
\end{array}\right. 
\end{equation}
 Here the amplitude $A_{\mu;e}$ can be identified as the LSND mixing
 amplitude $\sql$~\cite{LSND}.

 For the long-baseline reactor ($L\lsim 1$ km) 
 and accelerator experiments ($L< 1000$ km), 
 we have to include effects of the 
$\dma$ mass scale, and the conversion probability, 
$P_{\nu_\mu\to\nu_e}^\Ch$ becomes 
\begin{equation}
P_{\nu_\mu\to\nu_e}^\Ch = 1 - 2\, d_e(1-d_e) -    A_\Ch \, 
\sin^2 \left( \frac{\dma}{4E_\nu} L \right),
\end{equation}
with 
\begin{equation} \label{eq:defd1}
A_\Ch=
\left[\begin{array}{lr}
  4\,\eta_e(1-d_e-\eta_e) & 
\hskip 0.8cm \text{for {\it (3+1)} mass scheme,}\\ 
  4\,|U_{e2}|^2 |U_{e3}|^2 &
\hskip 1.2cm  \text{for {\it (2+2)} mass scheme.}
\end{array}\right. 
\end{equation}

 To accommodate the ({\it 2+2}) mass scheme, $d_\mu$, $A_\Ch$ and
 $A_{\mu;e}$ have to be small and $d_e$ close to 1 (to keep $A_{e;e}$
 amplitude small). Also the parameters $\eta_s$ and $d_s$ are not
 independent, $\eta_s=d_s $ as noted by the authors of
 Ref.~\cite{3+1}.  From Ref.\cite{valle-4nu} we know that the global
 fit of atmospheric and solar neutrino data prefers either the limit
 $\eta_s= d_s \sim 1 $ or the limit $\eta_s =d_s \ll 1 $. For the
 ({\it 3+1}) mass scheme, $d_\mu$,$d_e$, $A_\Ch$, $A_{\mu;e}$,
 $\eta_s$ and $1-d_s$, all, have to be small

 In this paper, in order to give an explicit example of the effect of
 four neutrino mass schemes we will fix the values of the mixing
 parameters.  These were taken within the parameter region allowed by
 solar and atmospheric neutrino data as well as by the null results of
 reactors and accelerators (with the exceptions of K2K~\cite{k2k} and
 KamLAND~\cite{kamland})~\cite{giunti,Gonzalez-Garcia:2002sm}.  We
 have scanned a grid of values to find the maximal possible values of
 the six mixing angles in order to maximize the oscillation effect 
 driven by the LSND mass scale in IceCube, using the constraints on 
 $d_\alpha$, $A_{\mu;e}$, $A_\Ch$, $\eta_e$ and $\eta_s$ shown in
 Ref.~\cite{valle-4nu}. The result of our scan is shown in
 Table~\ref{tab1}. For the neutrino mass splittings, we have used the
 best fit values $\dma=3\times 10^{-3}$ eV$^2$ and $\Delta
 m^2_{\myodot}=7\times 10^{-5}$ eV$^2$.  For definiteness we also have
 set $\vert \dml \vert =0.5$ eV$^2$.
\begin{figure}
\centering\leavevmode
\vglue -1.5cm
\hglue 1.0cm
\includegraphics[width=20cm]{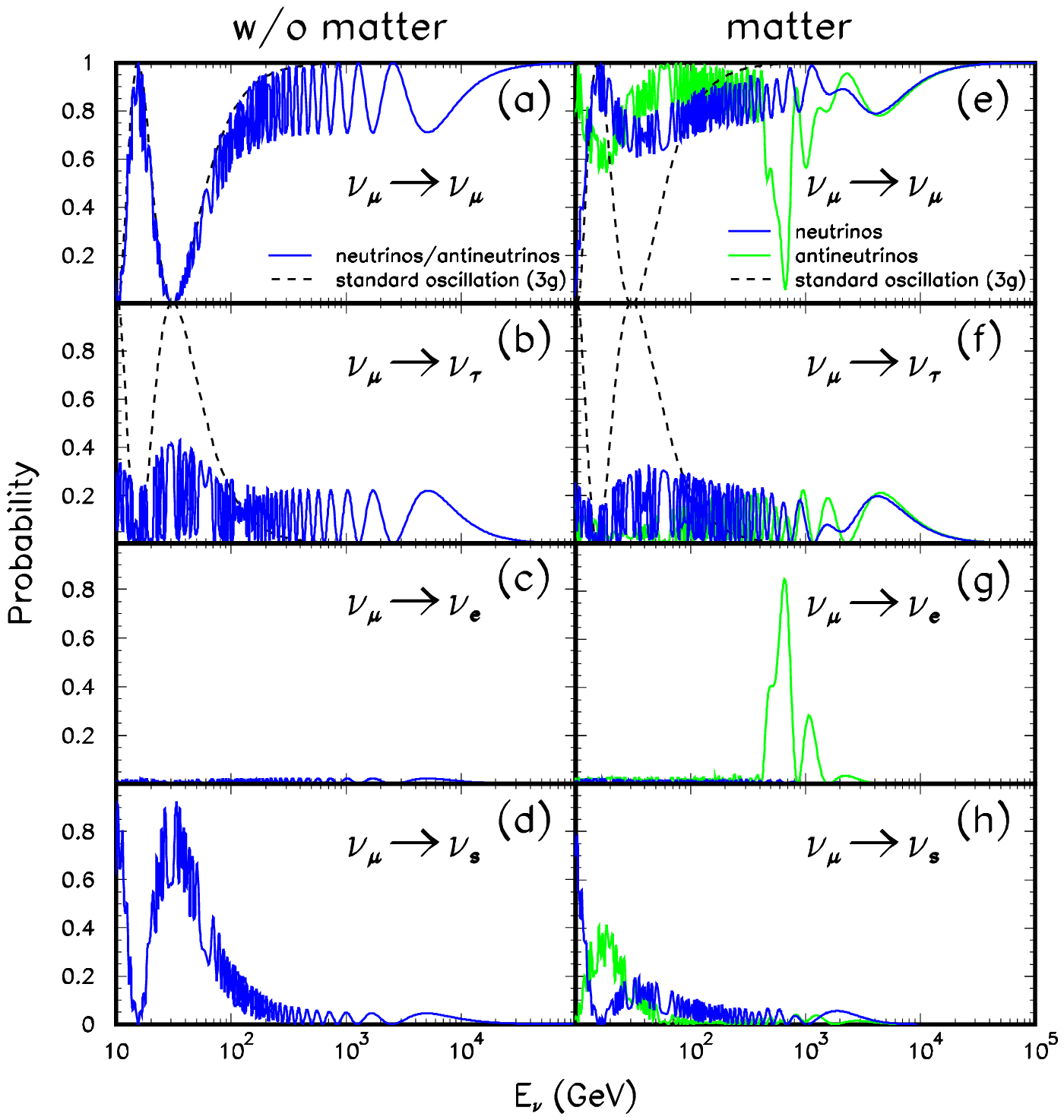}
\vglue -3.4cm
\caption{
 Neutrino and anti-neutrino probabilities, for diametral trajectories
($\cos \theta_z=-1$), in the {\em (2+2)} scheme
 with normal hierarchy for $\dml = 0.5$ eV$^2$
 {\it without} (left panels) and {\it with} (right panels) matter
 effects taken into account. We have used
 $\dma=3\times 10^{-3}$ eV$^2$, 
 $\dms=7\times 10^{-5}$eV$^2$ and the mixing
 parameter as given in Table~\protect\ref{tab1}. The standard three
generation oscillation probabilities are labeled as 3(g).}
\label{fig1a}
\end{figure}

 We have numerically solved Eq.~(\ref{evol}) to compute $P(\nu_{\mu}
 \rightarrow \nu_{\alpha})$ for $\alpha=e,\mu $,$\tau$ and $s$, the
 oscillation probabilities that are relevant for atmospheric neutrino
 data. Let us make some observations that can help the reader to understand
 our numerical results.  \\

 {\it (2+2) case:}\\
 
 From the expressions of the evolution
 equations, one can expect  matter enhanced MSW resonant conversion
 in Earth for the neutrino (antineutrino) channel $\nu_{\mu}
 \rightarrow \nu_e$ ($\bar{\nu}_{\mu} \rightarrow \bar{\nu}_e$), for
 the {\it 2+2} inverted (normal) hierarchy  if
 the following condition is satisfied,
\begin{equation}
 E_\nu^{\mathrm{R}} \displaystyle 
\sim \vert \dml \vert \times \frac {d_e - d_\mu}{2 A_{\rm{cc}}}
\sim 1.7
\mbox{ TeV} \times (d_e - d_\mu) \times 
\left(\frac{ \vert \dml \vert }{0.5 \, \mbox{eV}^2}\right) 
\times \left( 
\frac{2.0\, \mathrm{g/cc}}{Y_e \rho}\right).
\label{res}
\end{equation}
 Other resonances can be induced by the $\dma$ mass scale at $E_\nu
 \lsim 100$~GeV, but they are irrelevant for our purpose here.  In the
 {\it (2+2)} mass scheme, $d_e \sim 1$ and $d_\mu \ll 1$ so that the
 resonant condition can be achieved in the TeV energy range.  
 For the $\nu_\mu \rightarrow \nu_\tau$ channel we expect 
 almost vacuum oscillations.

\begin{table}
\begin{center}
\begin{tabular}{|c|c|c|c|c|c|c|c|c|}
\hline
Case & $\sin^2 \theta_{12}$ &$\sin^2 \theta_{13}$ &$\sin^2
\theta_{14}$ & $\sin^2 \theta_{34}$ & $\sin^2 \theta_{24}$ & 
$\sin^2 \theta_{23}$ & $\eta_s$ & $d_s$ \\
\hline
{\it 2+2} & 0.001 & 0.01 & 0.26 & 0.8 & 0.03 & 0.5 & 0.2 & 0.2  \\ 
{\it 3+1} & 0.26 & 0.035 & 0.02 & 0.45 & 0.045 & 0.5 & 0.13 & 0.52\\
\hline
\end{tabular}
\end{center}
\vglue -.5cm
\caption{
Values of the mixing parameters used in this paper. 
}
\label{tab1}
\end{table}

 In Fig.~\ref{fig1a} we show, for diametral trajectories
 ($\cos \theta_z=-1$), the behavior of the neutrino and antineutrino
 $\nu_\mu \to \nu_\mu$ and $\nu_\mu \to \nu_\tau, \nu_e, \nu_s$
 transition probabilities for the {\it (2+2)} case with normal
 hierarchy.  To provide a better understanding of what is going on, in
 panels (a)-(d) the matter effects were not taken into account while
 in panels (e)-(h) they were. We also show for comparison what would
 be expected for the standard three flavor neutrino oscillation
 scenario (labeled as 3(g)).

 If we compare Fig.~\ref{fig1a}(c) and Fig.~\ref{fig1a}(g), we see a
 sizable $\bar \nu_\mu \to \bar \nu_e$ conversion probability when
 matter effects are taken into account. This resonance behavior is
 due to the LSND mass scale and Earth matter effects~\cite{MSW}. Looking
 closely to Fig.~\ref{fig1a}(g), we can see three peaks: one at
 $E_\nu^{\mathrm{R}}\sim 0.57$ TeV for a typical value of Earth core
 density ($\rho_{\mathrm{core}}\sim 12 \mathrm{ g/cc}$), and another
 for $E_\nu^{\mathrm{R}}\sim 1.5$ TeV for a typical value of Earth
 mantle density, ($\rho_{\mathrm{mantle}}\sim 4 \mathrm{ g/cc}$). A
 third peak appears for $E_\nu^{\mathrm{R}}\sim 0.8$ TeV and its
 origin is due to a parametric resonance. Similar behavior was
 noticed sometime ago in the context of neutrino oscillations induced
 by $\dma$ for $\nu_{\mu} \rightarrow \nu_s$~\cite{smirnov-atm}.  To
 determine the behavior of oscillations in $\nu_\mu \to \nu_\tau$
 channel, we can use Eq.~(\ref{eq:pld}), changing $e\rightarrow \tau$,
 so that $P_{\nu_\mu\to\nu_\tau} = A_{\mu;\tau} \; \sin^2 (\dml L
 /(4E_\nu)) \,$. The conversion amplitude for this channel,
 $A_{\mu;\tau}= 4\cos^2 \theta_{34}
 \cos^2\theta_{24}\sin^2\theta_{24}$ depends only on
 $\sin^2\theta_{24}$ and $\sin^2\theta_{34}$. We confirmed numerically
 that this is correct and we have virtually vacuum oscillations
 induced by $\dml$ as you can see comparing Figs.~\ref{fig1a}(b) and
 (f).  In the case of inverted mass hierarchy, the neutrino and
 antineutrino probabilities are interchanged.\\

 {\it (3+1) case:}\\

 In the {\it (3+1)} mass scheme, we cannot expect a strong resonant 
 $\nu_\mu \to \nu_e$ ($\bar \nu_\mu \to \bar \nu_e$) transition due 
 to the LSND mass scale at TeV energies. This can be explained as follows. 
 First, both  $\nu_\mu$ and $\nu_e$ ($\bar \nu_\mu$ and $\bar \nu_e$) are 
 basically distributed among the mass triplet so that we cannot expect a
 significant direct transition between them driven by $\dml$.  
 However, a second order transition can be achieved via $\nu_4$ which 
 is mainly composed of $\nu_s$ as $\nu_\mu \to \nu_s \to \nu_e$ 
 ($\bar \nu_\mu \to \bar \nu_s \to \bar \nu_e$).

 For the normal mass hierarchy, the first step, $\nu_\mu \to \nu_s$ ($\bar
 \nu_\mu \to \bar \nu_s$), occurs via almost vacuum oscillation (a MSW
 resonance effect), while the second step, $\nu_s \to \nu_e$ ($\bar
 \nu_s \to \bar \nu_e$), occurs via a MSW resonance effect with 
  $E_\nu^{\mathrm{R}}$ given by Eq.~(\ref{res}) with $d_e-d_\mu$ replaced 
  by $d_s$ (almost vacuum oscillation).  
 The strength of these transitions is regulated by the value of 
 $\theta_{14}$ and  $\theta_{24}$, which are very much constrained by 
 current data (see Table I). This is why we do not expect to observe 
 a large number of $\nu_e$ ($\bar \nu_e$) events in the {\it (3+1)} scheme.   
 In the case of inverted mass hierarchy, the
 neutrino and antineutrino probabilities are interchanged.

  On the other hand, $\nu_\mu \to \nu_\tau$  
 ($\bar \nu_\mu \to \bar \nu_\tau$) transitions driven by the 
 LSND mass scale can occur directly mainly via vacuum-like oscillation, 
 since there is a significant amount of  $\nu_\tau$ in the $\nu_4$ state. 
 This is due to the large value of $\theta_{34}$ (see Table I), which is 
 the maximal value still allowed by data.
 
\begin{figure}
\centering\leavevmode
\vglue -1.5cm
\hglue 1.0cm
\includegraphics[width=20cm]{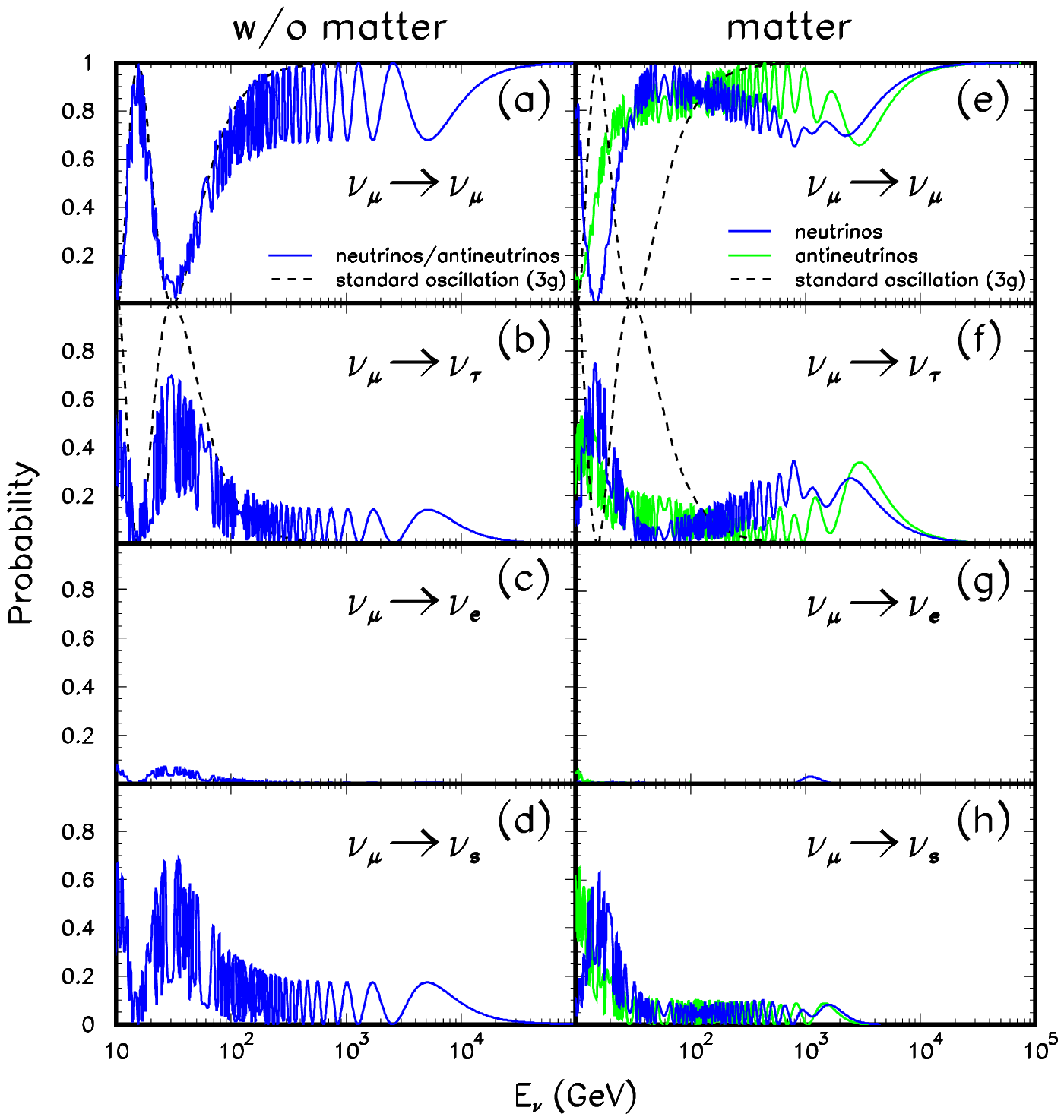}
\vglue -3.4cm
\caption{
 Neutrino and anti-neutrino probabilities, for diametral trajectories
($\cos \theta_z=-1$), in the {\em (3+1)} scheme
 with normal hierarchy for $\dml = 0.5$ eV$^2$
 {\it without} (left panels) and {\it with} (right panels) matter
 effects taken into account. We have used
 $\dma=3\times 10^{-3}$ eV$^2$, 
 $\dms=7\times 10^{-5}$eV$^2$ and the mixing
 parameter as given in Table~\protect\ref{tab1}. The standard three
generation oscillation probabilities are labeled as 3(g).}
\label{fig1b}
\end{figure}

 These observations are confirmed by our numerical calculations. 
 In Fig.~\ref{fig1b} we show, for diametral trajectories ($\cos
 \theta_z=-1$), the behavior of the neutrino and antineutrino $\nu_\mu
 \to \nu_\mu$ and $\nu_\mu \to \nu_\tau, \nu_e, \nu_s$ transition
 probabilities for the {\it (3+1)} case with normal hierarchy in a 
 similar way as in Fig.~\ref{fig1a}. \\ 

 From these observations we should expect at the TeV energy scale
 $\nu_\mu \to \nu_e$ matter enhanced resonant transitions for
 neutrinos (antineutrinos) if the LSND mass hierarchy is normal
 (inverted) in the {\it (2+2)} scheme and $\nu_\mu \to \nu_\tau$ as well as 
 $\bar \nu_\mu \to \bar \nu_\tau$ vacuum-like transitions in both mass schemes.

\section{Types of signal events}

 We have studied the observable consequences of these four neutrino
 schemes at high energies, in a future neutrino telescope.  For
 definiteness we have assumed a kilometer-scale neutrino telescope as
 the one proposed by the IceCube collaboration~\cite{icecube} that,
 for simplicity, has been simulated with perfect efficiency as well as
 perfect angular and energy resolution.  We have supposed that the
 IceCube detector can be described by a cylinder with 900 m of height
 and 500~m radius.

  We have calculated two types of measurable signals in this detector:
  upward-going muons and cascades events.  The calculations were done
  using the atmospheric neutrino fluxes, denoted by
  $\Phi_{\nu_\alpha}$, $\alpha=e,\mu$, of the Bartol
  group~\cite{bartol} and the cross sections from Ref.~\cite{gandhi}.
  We have also included in our calculations the effect of neutrino
  absorption when crossing the Earth, taking from Ref.~\cite{gandhi}.

\subsection{Upward-going muons}

 Muon neutrino charged current interactions, occurring inside the
 Earth close enough to the detector, 
 may produce energetic muon events which can cross the
 detector volume, giving rise to the so-called upward-going
 muons. These events are characterized by a very good angular
 resolution.

 We have calculated the number of upward-going muon events as a
 function of the zenith angle, $N_\mu(\theta_z)$, as in
 Refs.~\cite{lipari,nos},
\begin{equation}
N_\mu(\theta_z)=T \int_{E_{\mu,\mathrm{min}}}^{\infty}
A(R_{\mathrm{min}},\theta_z)
\frac{d\Phi_\mu(E_\mu,\cos\theta_z)}{dE_\mu d\cos\theta_z}dE_\mu + \text{antineutrinos}, \, \; 
\label{upmuons1}
\end{equation}  
with
\begin{eqnarray}
\frac{d\Phi_\mu}{dE_\mu d\cos\theta_z}
&=& N_A \int_{E_{\mu}}^\infty dE_{\mu 0} 
\int_{E_{\mu 0}}^\infty dE_\nu
 \int_0^\infty dh \;
\kappa_{\nu_\mu}(h,\cos\theta_z,E_\nu) \nonumber \\
& & \times \frac{d\Phi_{\nu_\mu}(E_\nu,\theta_z)}{dE_\nu d\cos\theta_z}
P (\nu_\mu \to \nu_\mu )
\frac{d\sigma(E_\nu,E_{\mu 0})}{dE_{\mu 0}}\,
F_{\mathrm{rock}}(E_{\mu 0}, E_\mu) S(E_\nu,\theta_z),
\label{upmuons2}
\end{eqnarray}
 where $T$ is the livetime that we assumed to be 1 yr, $N_A$ is the
 Avogadro number, $E_{\mu 0}$ is the energy of the muon produced in
 the neutrino interaction and $E_\mu$ is the muon energy when entering
 the detector, $F_{\mathrm{rock}}(E_{\mu 0}, E_\mu)$ is a function
 that takes into account the muon energy loss from the point where the
 muon is created to where it is detected~\cite{lipari}, $\kappa_\alpha$
 is the slant distance distribution normalized to
 one~\cite{pathlength}, $P (\nu_\mu \to \nu_\mu )$ is the muon
 survival probability, $A(R_{\mathrm{min}},\theta_z)$ is the effective
 area for a muon traveling a distance $R_{\mathrm{min}}$ inside the
 detector coming from the direction $\theta_z$~\cite{lipari}, $\sigma$
 is the $\nu_\mu N \to \mu X$, neutrino cross section, $\cos\theta_z$
 labels both the neutrino and the muon directions which to a very good
 approximation at the relevant energies are collinear,
 $S(E_\nu,\theta_z)$ is a shadow factor that is related to the
 attenuation of neutrinos traversing the Earth~\cite{gandhi}.  For a
 given zenith angle, the threshold energy cut is obtained by equating
 $R(E_{\mu,\rm{min}}) =R_{\mathrm{min}}=100$ m, where
 $R(E_{\mu,\rm{min}})$ is the muon range function.

\begin{figure}
\centering\leavevmode
\vglue -3.0cm
\hglue 1.8cm
\includegraphics[width=20.cm]{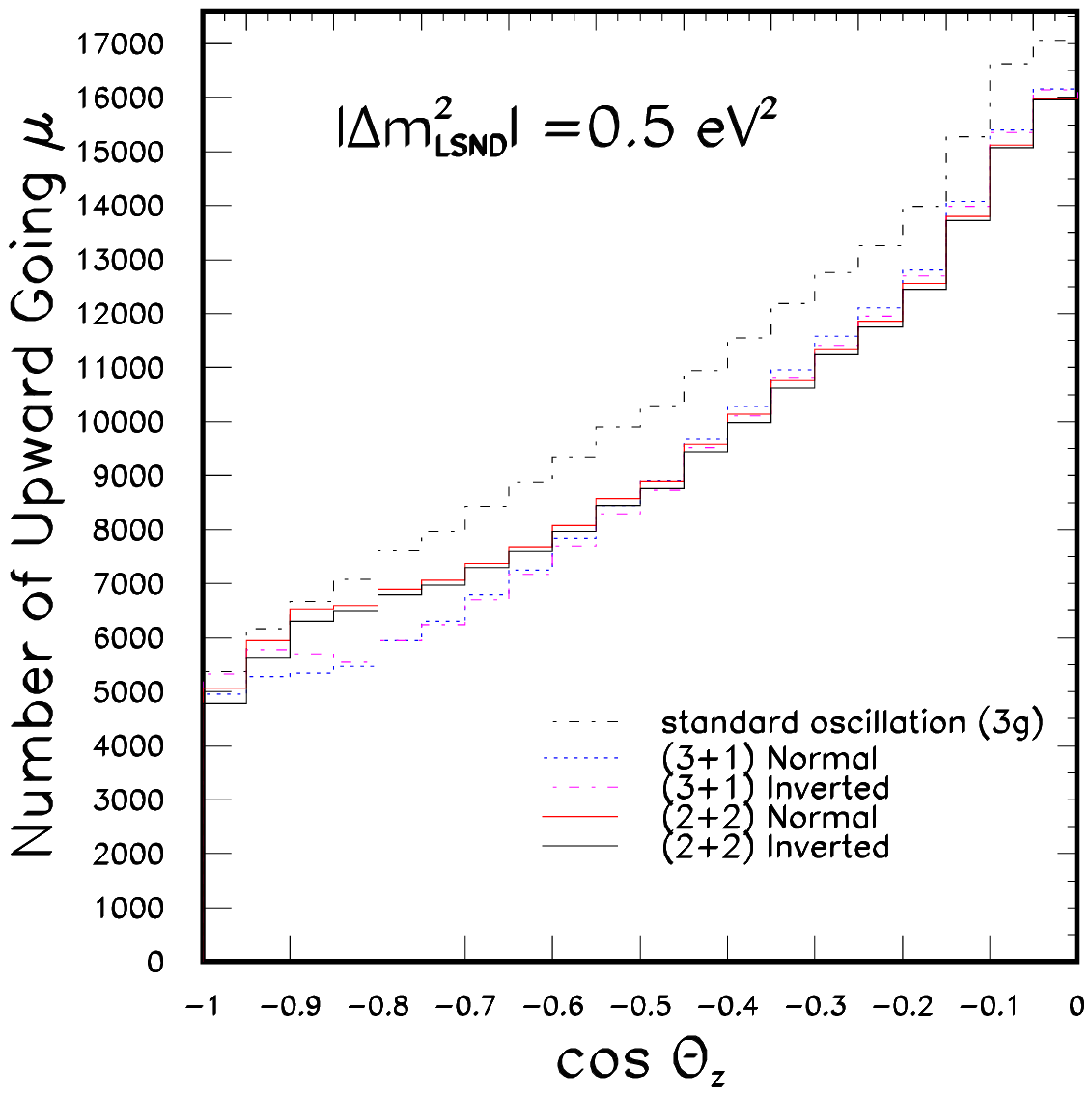}
\vglue -5.0cm
\caption{Expected zenithal distribution of  upward-going muon events in the 
 IceCube experiment after one year of data taking for the standard 
 three flavor oscillations (3g) and the different four neutrino 
 mass schemes.}
\label{fig2}
\end{figure}

 In Fig.~\ref{fig2} we show the zenithal angle distribution of these
 events expected in the case of the standard three flavor oscillations
 (or no oscillations at all) and for each of the four neutrino mass
 schemes assuming $\vert \dml \vert=0.5$ eV$^2$ and the oscillation
 parameters as given in Table~\ref{tab1}.  It is clear from this plot
 that the distribution of upward-going muons reflects the
 $\nu_\mu/\bar \nu_\mu$ disappearance into other flavors promoted by
 each of the four neutrino schemes, making them quite distinguishable
 from the standard three flavor oscillation case. It may be even
 possible to distinguish among {\it (2+2)} and {\it (3+1)} schemes,
 although their normal and inverted cases are virtually the same due
 to the fact that the mean energy of these events is much lower than
 the resonant energy ($E_\nu^{\mathrm{R}} \sim 0.8$ TeV for $\dml =
 0.5$ eV$^2$).
 From the Eq.~(\ref{upmuons1}) we note that all muons with an energy
 sufficient high to travel a distance $R_{\mathrm{min}}$ are classified
 as upward-going muons, the original muon energy is integrated out so that 
 the correlation between the original neutrino energy and the final muon 
 energy is weak. 

 As a test of capability to discriminate the four neutrino mass
 schemes from the standard three generation oscillation we performed a
 simple $\chi^2$ analysis, as follows. If we define $N_\mu^i(4\nu)$
 ($N_\mu^i$(3g)) as the number of upward-going muons in the $i$-th 
 zenith angle bin in four neutrino mass schemes (standard three
 neutrino mass scheme) showed in Fig.~\ref{fig2}, then we can compute
 the function 
\begin{equation}
\chi^2=\sum_{i=1,20}\frac{(\eta \; N_\mu^i(4\nu)-N_\mu^i(3g))^2}{(\sigma_{(4\nu)}^i+\sigma_{(3g)}^i)^2},
 \end{equation} where $\sigma_{(4\nu)}^i$ ($\sigma_{(3g)}^i$) is the
 statistical error on $N_\mu^i(4\nu)$ ($N_\mu^i(3g)$), for simplicity
 we have ignored systematic uncertainties. The parameter $\eta$ is a
 normalization factor. If we assumed predictions of upward going muons
 in standard three generation are the correct ones, then the only
 possible change is a global normalization factor. We scan the
 parameter $\eta$ to look for the best fit and for any of neutrino
 mass schemes the minimum $\chi^2$ is much higher then the degrees of
 freedom, 20-1 (20 points minus 1 normalization factor),
 with a goodness of fit well below $1\%$. Then we conclude that for
 the choice of mixing parameters in this paper, it is very likely that
 the zenith angular distribution of upward going muons is sufficient
 to be separated from the usual standard three neutrino
 oscillations. From this we can say that independent of which four
 neutrino mass scheme nature chooses, we can always have a distinctive
 zenith distortion in upward-going muons.

\subsection{Cascades}

  Electromagnetic and hadronic showers can be produced by ionizing
  products of the neutrino interactions with the ice. These types of
  events can be created by electron neutrino charged current
  interactions ($\nu_e-\rm{CC}$), tau neutrino charged current
  interactions ($\nu_\tau-\rm{CC}$) and neutral current interactions
  ($\nu-\rm{NC}$). These events, which are called cascade events, are
  characterized by a very good energy resolution. In principle, muon
  neutrinos also produce a hadronic shower and a muon and this case
  contribute to cascade events. We assume events with a muon track
  always can be separated from the other cascade events and we will
  not include them in our computation.  We have calculated the number
  of cascade events for a given zenith angle,
  $N_{\mathrm{cascade}}(\theta_z)$, as in
  Refs.~\cite{stanev,reno},
\begin{equation}
N_{\mathrm{cascade}}(\theta_z)=T V N_A \int_{E_{\mathrm{shr,min}}}^{\infty}
\frac{d\Phi_{\nu_\mu}(E_{\nu},\cos\theta_z)}{dE_{\nu}
d\cos\theta_z}
P (\nu_\mu \to \nu_e ) \sum_i f_i(E_\nu)dE_{\nu} + \text{antineutrinos},
\label{cascade1}
\end{equation}  
 here $V$ is the IceCube experiment volume (1 km$^3$), $f_i(E_\nu,z)$,
$i=\nu_e-$CC, $\nu_\tau-$CC (hadronic plus electromagnetic showers),
$\nu-$NC, are  functions which depend on the  interaction,
\begin{equation}
\begin{array}{lr}
 f_{\nu_e-\rm{CC}}(E_\nu) = \displaystyle \int dy \,
\frac{d\sigma_{cc}(E_\nu,y)}{dy} \,
\Theta(E_\nu-E_{{\mathrm{shr,min}}}), &  \hskip 1.2cm \text{for} \; \nu_e-\text{CC} \\
 f_{\nu_\tau-\rm{CC}}^1(E_\nu) = \displaystyle \int dy \int dz \,
 \frac{d\sigma_{cc}(E_\nu,y)}{dy} \, \frac{dn(E_\tau)}{dz}
\Theta(E_\nu (y+(1-y)(1-z))-E_{\mathrm{shr,min}}), &  \hskip 1.2cm \text{for}
\; \nu_\tau-\text{CC hadronic}\\
 f_{\nu_\tau-\rm{CC}}^2(E_\nu) = \displaystyle \int dy \int dz^ {\prime} \,
\frac{d\sigma_{cc}(E_\nu,y)}{dy} \, \frac{dn(E_\tau)}{dz^{\prime}} 
\Theta(E_\nu (y+(1-y)z^{\prime})-E_{\mathrm{shr,min}}), &  \hskip 1.2cm
\text{for} \; \nu_\tau-\text{CC electromagnetic} \\
 f_{\nu-\rm{NC}}(E_\nu) = \displaystyle \int dy \,
\frac{d\sigma_{nc}(E_\nu,y)}{dy} \, 
\Theta(E_\nu y-E_{\mathrm{shr,min}}),  &  \hskip 1.2cm
\text{for} \; \nu-\text{NC}  \\ \nonumber
\end{array}
\label{cascade2}
\end{equation}
 where $E_\tau$ is the tau energy, $y,z,z^\prime$ are the original
 neutrino energy fractions carried by the relevant lepton/hadron at
 play, and we have considered only events which have a total shower
 energy greater than 500 GeV ($E_{\mathrm{shr,min}}=500$ GeV).

 For $\nu_\tau-$CC events we can have hadronic and electromagnetic
 cascades, respectively, from $\nu_\tau \to \tau \to e $ decay channel
 and from $\nu_\tau \to \tau \to h$, where $h$ means any hadrons. For
 $\nu-$NC ($\nu_e-$CC) events we have respectively hadronic
 (electromagnetic) showers.  The functions $dn (E_\tau)/dz$ and $dn
 (E_\tau)/dz^{\prime}$ are given in Ref.~\cite{reno}. Although we also
 integrate on the neutrino energy as for the upward-going muons
 computation, the average shower energy for this sample is more
 closely related to the average initial neutrino energy.

  Since for the energies considered here ($E_{\nu} \gsim 1$ TeV), the
  atmospheric neutrino fluxes are basically composed of $\nu_{\mu}$
  and $\bar{\nu}_{\mu}$~\cite{review-atm} and according to the usual
  three neutrinos oscillation scenario electron and tau neutrino
  appearance are suppressed, therefore one generally expects that only
  the neutral current contribution can produce cascades at these
  energies.  A possible exception to this would be perhaps charged current
  contributions induced by $\nu_e$/$\nu_\tau$ neutrinos emitted in
  microquasar jets~\cite{gdlw}, but these events are very well
  collimated with the source and can be easily separated from the
  atmospheric neutrino data.  As we have shown earlier, in four
  neutrino schemes, electron and tau neutrino components will appear
  at these energies so that an increase on the number of cascade
  events should be expected.

\begin{figure}
\centering\leavevmode
\vglue -3.0cm
\hglue 1.8cm
\includegraphics[width=20.cm]{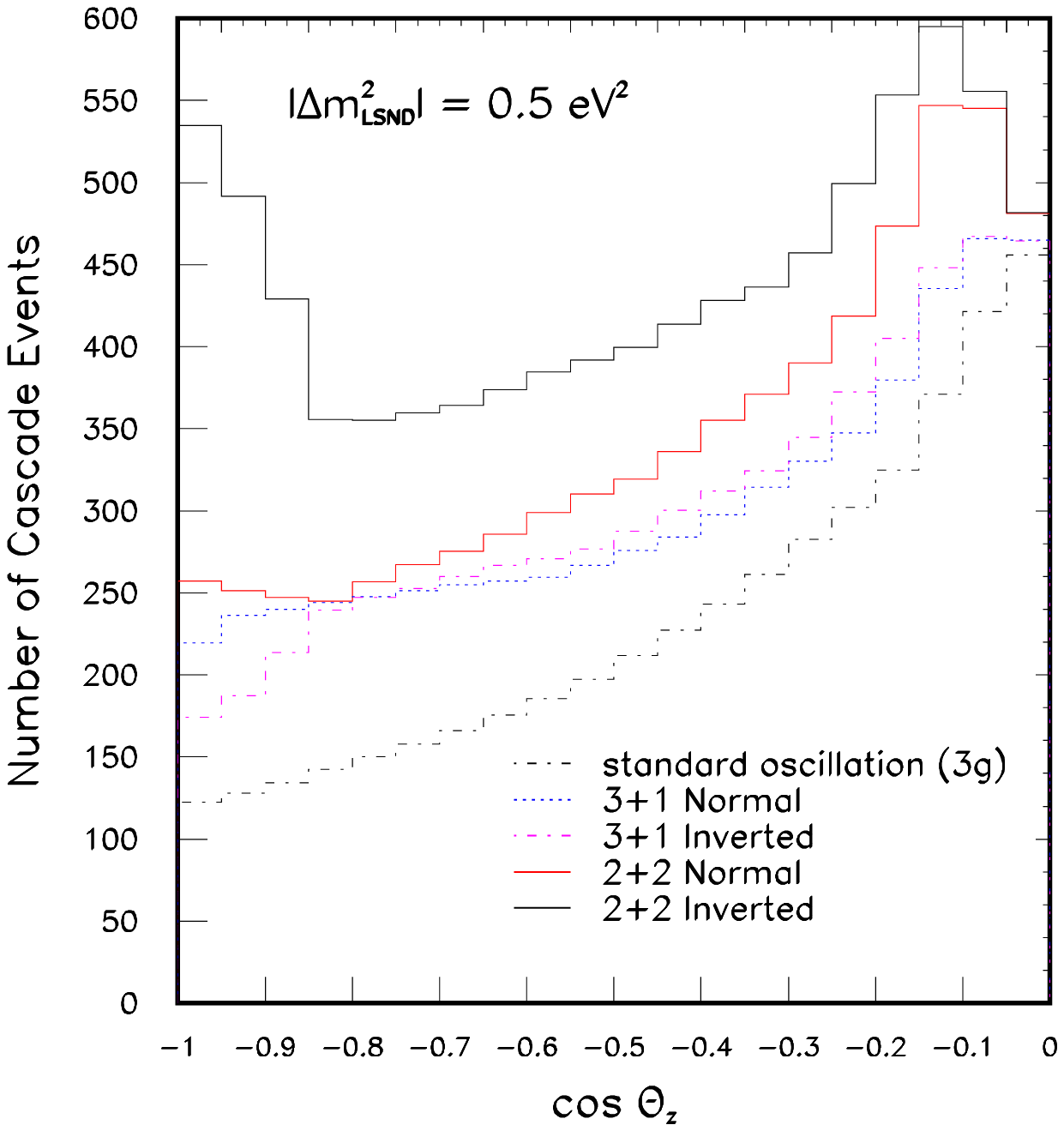}
\vglue -4.0cm
\caption{Expected zenithal distribution of cascade events in the
 IceCube experiment after one year of data taking for the standard 
 three flavor oscillations (3g) and the different four neutrino 
 mass schemes.}
\label{fig3}
\end{figure}

 We show in Fig.~\ref{fig3} the zenithal distribution of cascade
 events that can be expected in IceCube after one year of data taking,
 for the standard three neutrino oscillation case as well as for all
 the four neutrino schemes, assuming $\vert\dml\vert=0.5$
 eV$^2$ and the oscillation parameters as given in Table~\ref{tab1}.
 All four neutrino cases, for these values of the parameters, 
 can be distinguished from the standard case.
 We observe that there is an increase on the number of cascade events
 at $\cos \theta_z \sim -1$ for the {\it (2+2)} inverted scheme due to
 the resonant appearance of $\nu_e$ similar to what is shown in
 Fig.~\ref{fig1a} (g) for $\bar \nu_e$.  There is no such a prominent peak
 for the {\it (2+2)} normal scheme due to the fact that in this case
 the resonance is in the antineutrino channel (where one has a lower
 flux and smaller cross section).  The increase of the number of
 cascade events in the {\it (3+1)} schemes is almost exclusively due
 to $\nu_\tau$ appearance.  In fact, the angular distribution of
 cascade events reflect the flavor content of the atmospheric neutrino
 flux as measured at the detector site.

 We observe that cascade events are more effective in separating the
 mass hierarchy of four neutrino schemes than the upward-going muon
 events. This is not surprising since the 500 GeV energy cut imposed
 on the cascade showers selects events closer to the resonant energy
 where the effect of the large $\vert \dml\vert =0.5$ eV$^2$ is more
 sizable.  For higher values of $\vert \dml \vert $ the differences
 among the standard and the four neutrino oscillations distributions
 could be even larger than what we have shown in Figs.~\ref{fig2} and
 \ref{fig3} if not for the fact the mixing angles are much more
 constrained by data for higher masses~\cite{valle-4nu}.  In any case
 the general behavior of the curves shown in this paper will still
 hold for $\vert \dml \vert $ as high as 2.0 eV$^2$.

 We note that unlike the case of upward-going muon events,
 cascade events can be observed also for $\cos \theta_z >0$, which
 correspond to down-going events. For such events, for any four
 neutrino mass scheme we have considered here, we do not expect any
 significant deviation of the zenith angle distribution from what is
 expected in the standard three neutrino oscillation scheme. In the
 standard scheme, in good approximation, the zenith angle distribution
 is symmetric with respect to $\cos \theta_z=0$. This is because of
 the fact that for down-going cascade events, there is no matter
 effect and the distance traveled by neutrinos is much smaller than
 the typical oscillation length expected for $\vert \dml \vert$ scale
 and $E_\nu \sim 1$ TeV so that no significant $\nu_e,\bar
 \nu_e/\nu_\tau, \bar \nu_\tau$ induced events are expected. Therefore,
 by comparing the zenith distributions for $\cos \theta_z>0$ and $\cos
 \theta_z<0$, it is expected that we can discriminate four neutrino
 mass schemes from the standard three one with better sensitivity.

 Finally, let us note that the uncertainty in the prediction of the
 atmospheric muon neutrino flux, which is currently about 15\%, could
 potentially spoil the discriminating power among the different
 signals. This can be, to some extent, overcome by combining cascade
 and upward-going muon events, and so reducing the uncertainty in the
 overall flux.  Also, the shape of the event distributions, here the
 key point in disentangling the different signals, is not very much
 affected by the overall flux normalization.

\section{Conclusions}

 We have studied the r\^ole that can be played by a large mass
 splitting implied by the LSND signal, $\vert \dml \vert \sim
 (0.5-2.0)$ eV$^2$ in a four neutrino mass scheme for atmospheric
 neutrino data in the energy range $\sim$ 1 TeV.  We have shown that
 upward-going muon events as well as cascade events, that can be
 measured by a kilometer scale detector such as the one proposed by
 the IceCube collaboration, can indicate the presence of the fourth
 neutrino through the production of $\nu_e,\bar \nu_e$ and
 $\nu_\tau,\bar \nu_\tau$ which are not expected in the standard three
 generation oscillation case where we have only the small mass splittings 
 $\dms$ and $\dma$.
 Furthermore the zenithal dependence of cascade events is even
 sensitive to the sign of $\dml$ because the matter effect play an
 important r\^ole, leading to resonant $\nu_\mu \to \nu_e$ or $\bar
 \nu_\mu \to \bar \nu_e$ conversion depending on the sign of $\dml$.
 We note that the standard three generation oscillation scheme cannot
 induce these signals of $\nu_e$/$\nu_\tau$ appearance at the 1 TeV
 scale, and therefore, they constitute an unique signature of a large
 mass scale present in a four neutrino mass scheme.
 
 A final remark. If MiniBooNE observes a positive signal, the effects 
 shown here should be expected independent of the four neutrino mass 
 scheme. On the other hand, if MiniBooNE results turns out to be negative
 this may be viewed  as a complementary test of the presence of a large 
 neutrino mass splitting in nature.

\begin{acknowledgments} 
  This work was supported by Funda\c{c}\~ao de Amparo
 \`a Pesquisa do Estado de S\~ao Paulo (FAPESP), Conselho
 Nacional de Ci\^encia e Tecnologia (CNPq), Fundo de Apoio ao Ensino e
\`a Pesquisa (FAEP) and by the National Science
 Foundation under Grant No. PHY99-07949.  We thank T. Stanev for
 providing the table of the atmospheric neutrino fluxes. We also thank
 C.~Lunardini for bringing to our attention Ref.~\cite{smirnov-atm}
 and A. de Gouvea for useful comments. We are grateful for the
 hospitality of the Kavli Institute for Theoretical Physics, of the
 University of California in Santa Barbara, where this work was
 completed.
\end{acknowledgments}


\end{document}